\documentstyle[11pt,aaspp4]{article}

\begin{document}

\input{psfig.tex}

\title{ASCA Observations of Radio-Loud AGNs}

\author{Rita M. Sambruna \& Michael Eracleous}
\affil{Department of Astronomy \& Astrophysics, 
The Pennsylvania State University \\
525 Davey Lab, University Park, PA 16802}

\author{and}

\author{Richard F. Mushotzky}
\affil{NASA/GSFC, Code 662, Greenbelt, MD 20771}

\begin{abstract}

We summarize the results of an X-ray spectroscopic survey of radio-loud AGNs
observed with {\it ASCA}, using proprietary and archival data (public up to
1998 September). We briefly compare our results with those obtained
for radio-quiet AGNs studied by other authors, and with the predictions
of unified models.

\end{abstract}

\section{Introduction}

One of the fundamental open problems of AGN research is the dichotomy
between radio-loud and radio-quiet (RL/RQ) AGNs. While in both cases
the ultimate source of power is thought to be accretion onto a massive
black hole (Antonucci 1993; Urry \& Padovani 1995), the two classes
exhibit subtle but systematic differences at all observed wavelengths,
the most conspicuous being the ability of RL AGNs to form collimated,
relativistic jets, in contrast to their RQ counterparts.

The X-ray spectra of AGNs may hold the key to understanding the RL/RQ
dichotomy.  Since X-rays are produced in the innermost regions of the
accretion flow, they allow us a direct view of the central engine
where the jets form.  Indeed, different models for the formation of
jets require different structures for the central engines and thus
make different predictions for the production of X-rays, e.g., the
spinning black hole model of Blandford \& Znajek (1977) combined with
the hydromagnetic wind model of Blandford \& Payne (1982), or the ion
torus or advection-dominated disk of Rees et al. (1982) and Narayan \&
Yi (1994,1995). 
It is thus encouraging that previous studies of RL AGNs with the {\it
Einstein} IPC and EXOSAT found systematic differences, with RL sources
being more X-ray-luminous and having flatter X-ray spectra
($\Delta\Gamma \sim 0.5$) than RQ ones (Wilkes \& Elvis 1987; Lawson
et al. 1992; Shastri et al. 1993; Lawson \& Turner 1997). However,
these studies were plagued by the low sensitivity of the instruments,
especially at soft energies where the contribution of diffuse thermal
components can be present (Worrall et al. 1994).

We started a systematic study of RL AGNs in the X-ray band using {\it
ASCA} archival data and new observations. The improved sensitivity and
resolution of {\it ASCA} allows us to study in great detail the
0.6--10~keV spectra, especially in the region of the Fe K$\alpha$
line, disentangling the AGN emission from any thermal contribution.
While our principal goal is to compare the X-ray properties of RL and
RQ AGNs, we will also test the idea that different RL subclasses
differ by orientation only, as postulated by current RL unification
scenarios.  We report here our preliminary results, referring the
reader to Sambruna, Eracleous, \& Mushotzky (1999, in prep.) for a
complete description of our work and our conclusions.

Our sample contains 37 RL AGNs, including 10 Broad Line Radio Galaxies
(BLRGs), 6 RL Quasars (QSRs), 12 Narrow-Line Radio Galaxies (NLRGs),
and 10 Radio Galaxies (RGs). The classification was based on the
luminosity of the [O{\,\sc iii}] emission line, with type-1 BLRGs and
QSRs having $\log L_{\rm [O{\,\sc iii}]} < 43.7$ and $> 43.7$ erg
s$^{-1}$, respectively, and type-2 NLRGs and RGs having $\log L_{\rm
[O{\,\sc iii}]} > 41.0$ and $< 41.0$ erg s$^{-1}$, respectively.
The sample is by no means statistically complete, and it probably
reflects a bias toward the brightest sources of each type.  The
detection rate with {\it ASCA} is high: 100\% BLRGs, 83\% NLRGs, 83\% QSRs,
and 90\% RGs were detected by both the SIS and GIS, but the
signal-to-noise ratio in the SIS has a wide range of 3--300.

\begin{figure}
\noindent{\psfig{figure=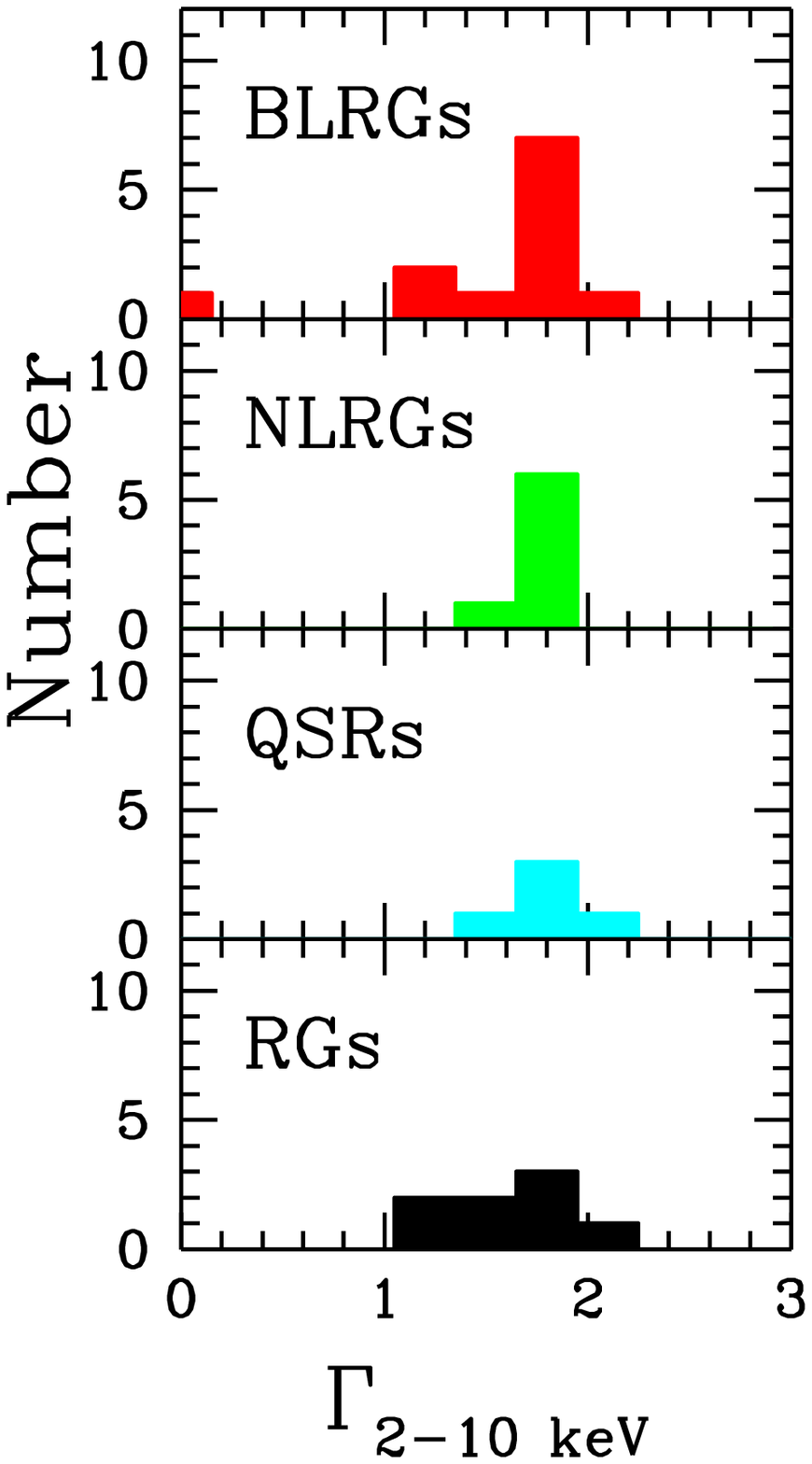,height=2.7in}}{\psfig{figure=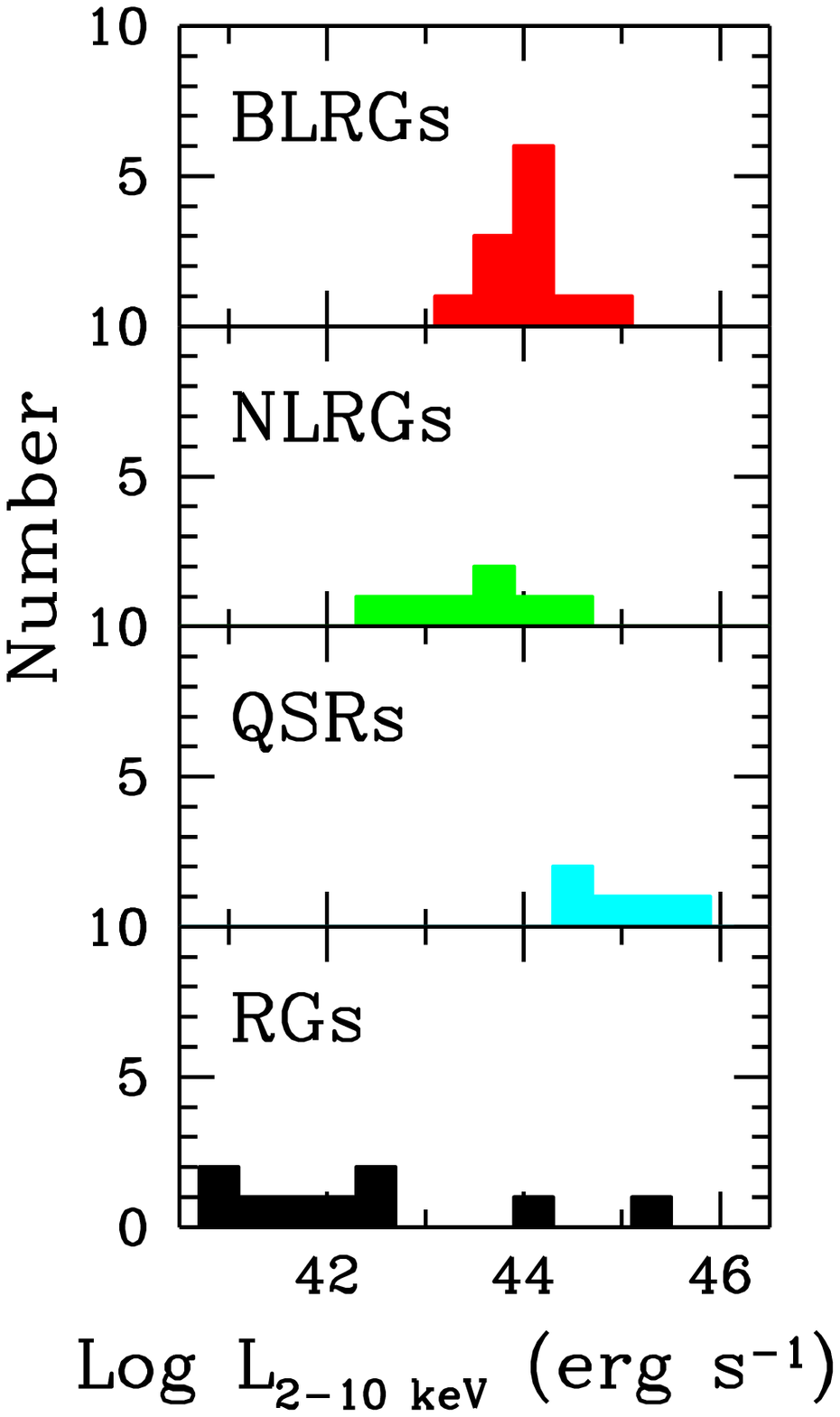,height=2.7in}}{\psfig{figure=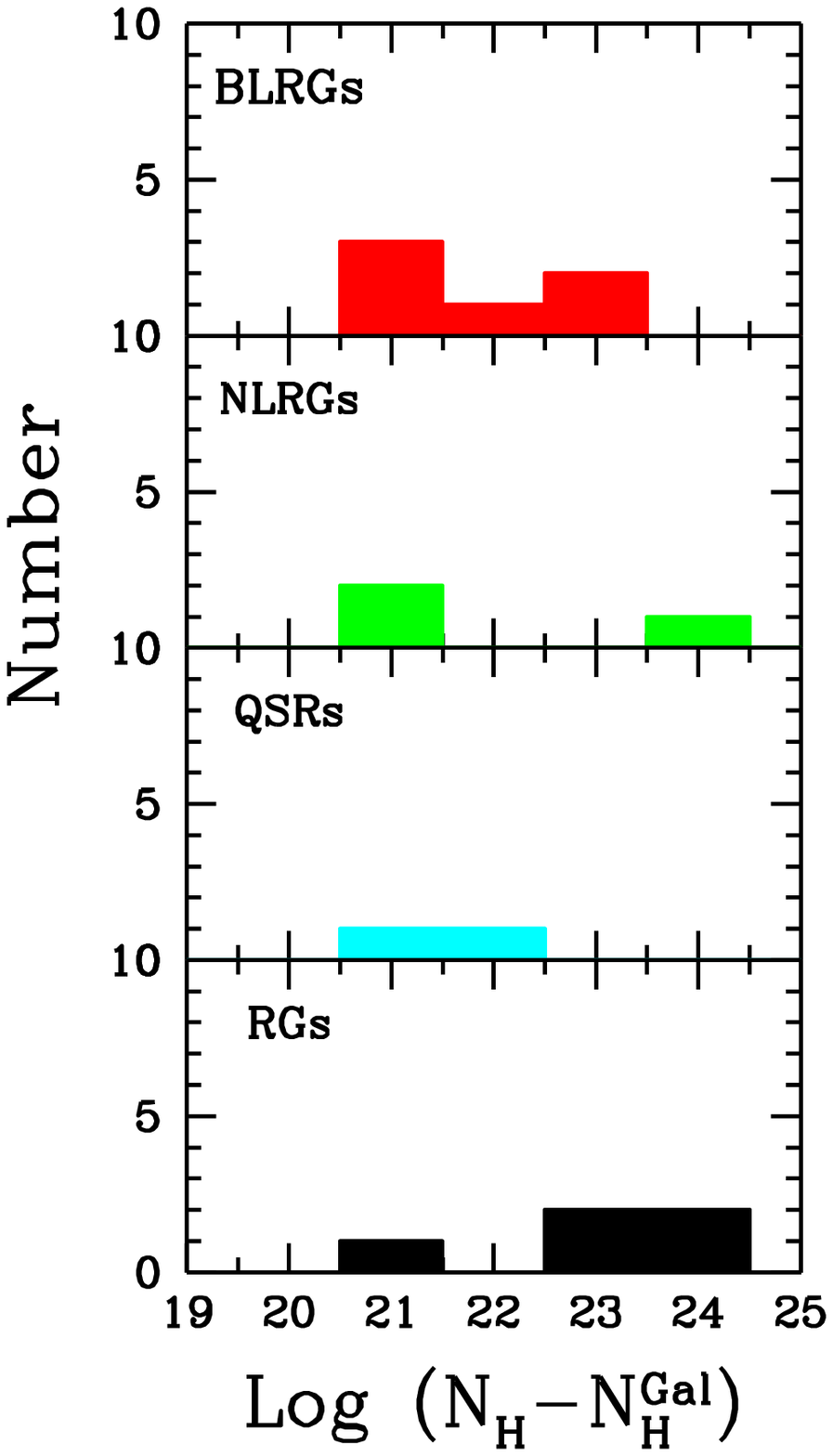,height=2.7in}}
\caption{The distributions of the 2--10~keV photon index {\it (a,
Left)}, absorption-corrected 2--10~keV luminosity {\it (b, Middle)},
and of the difference of the fitted and Galactic column densities {\it
(c, Right)}, excluding upper limits.}
\end{figure}


\section{Summary of ASCA Results} 

\begin{enumerate}

\item
At soft energies, thermal emission is present in 40\% of type-2
sources with temperatures and luminosities typical of poor groups or
clusters. In two cases the 0.5--4.5~keV intrinsic luminosity of the
thermal component is consistent with the predicted luminosity of a
starburst.

\item 
A thermal soft excess is detected in one BLRG below 2 keV, best
modelled with a $kT \sim 2$ keV bremmsstrahlung component. 

\item
A power law component is detected in 92\% of the sources (34/37) at
energies above 2~keV, with a narrow distribution of photon indices
$\Gamma$ for the various subclasses (Figure~1$a$). The average photon
index $\Gamma$ is similar, $\langle \Gamma \rangle \sim 1.7-1.8$.
 
\item
NLRGs and RGs have wide distributions of intrinsic
(absorption-corrected) 2--10~keV luminosities, extending more than one
order of magnitude below those of BLRGs and QSRs (Figure~1$b$).

\item
Excess cold absorption over Galactic of the hard power law component
is detected in 64\% NLRGs and 100\% RGs (Figure~1$c$), with $N_{\rm H}
\sim 10^{21-23}$ cm$^{-2}$. Interestingly, a few BLRGs and QSRs
exhibit similar excess columns as well.

\item
Absorption edges between 0.7--1~keV (the trademark of a warm absorber)
are detected in only one BLRG, with optical depth $\tau \sim 0.3$.
 
\item
The Fe K$\alpha$ line is detected in 50\% BLRGs, 33\% NLRGs, 20\%
QSRs, and 22\% RGs. In most cases the line is unresolved; in a few
BLRGs the line is broad (Gaussian $\sigma \sim 0.3-0.5$~keV).
 
\item
The intrinsic 2--10~keV power law luminosity is strongly correlated
with the luminosity of the [O{\,\sc iii}] emission line (Figure~2$a$);
a partial correlation test (taking into account the non-detections)
gives a Kendall correlation probability $P_{\rm K} \sim$ 99.3\% after
the redshift dependence is removed. The X-ray luminosity is also
weakly correlated with the far-infrared (FIR) luminosity at 60$\mu$
(Figure~2$b$), $P_{\rm K} \sim$ 91\%. No significant correlation is
found from the partial analysis between the 2--10~keV luminosity and
the extended radio luminosity at 5 GHz or the ratio of the radio
core-to-lobe power.
 
\end{enumerate}


\begin{figure}
\noindent{\psfig{figure=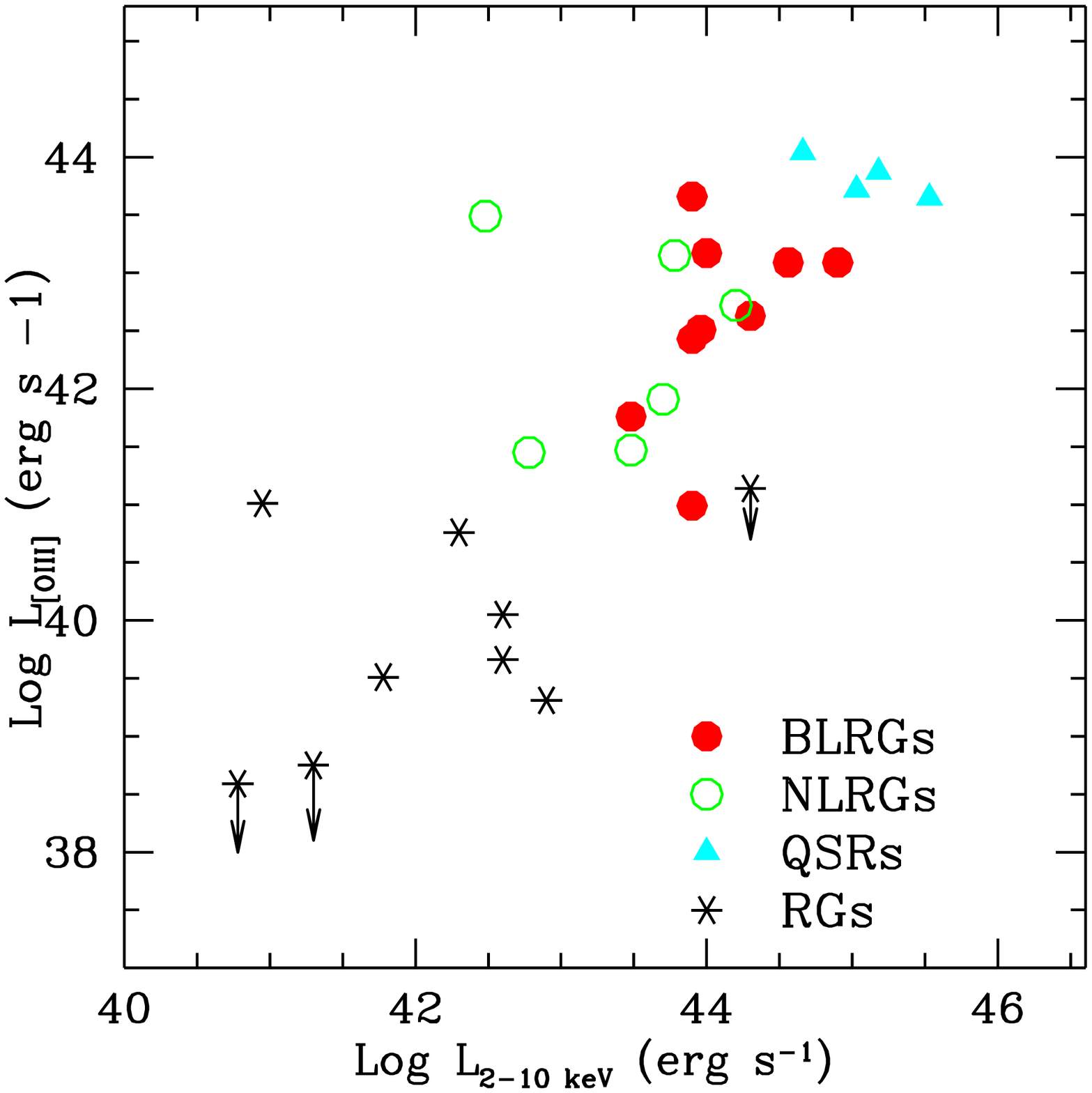,height=3.5in}}{\psfig{figure=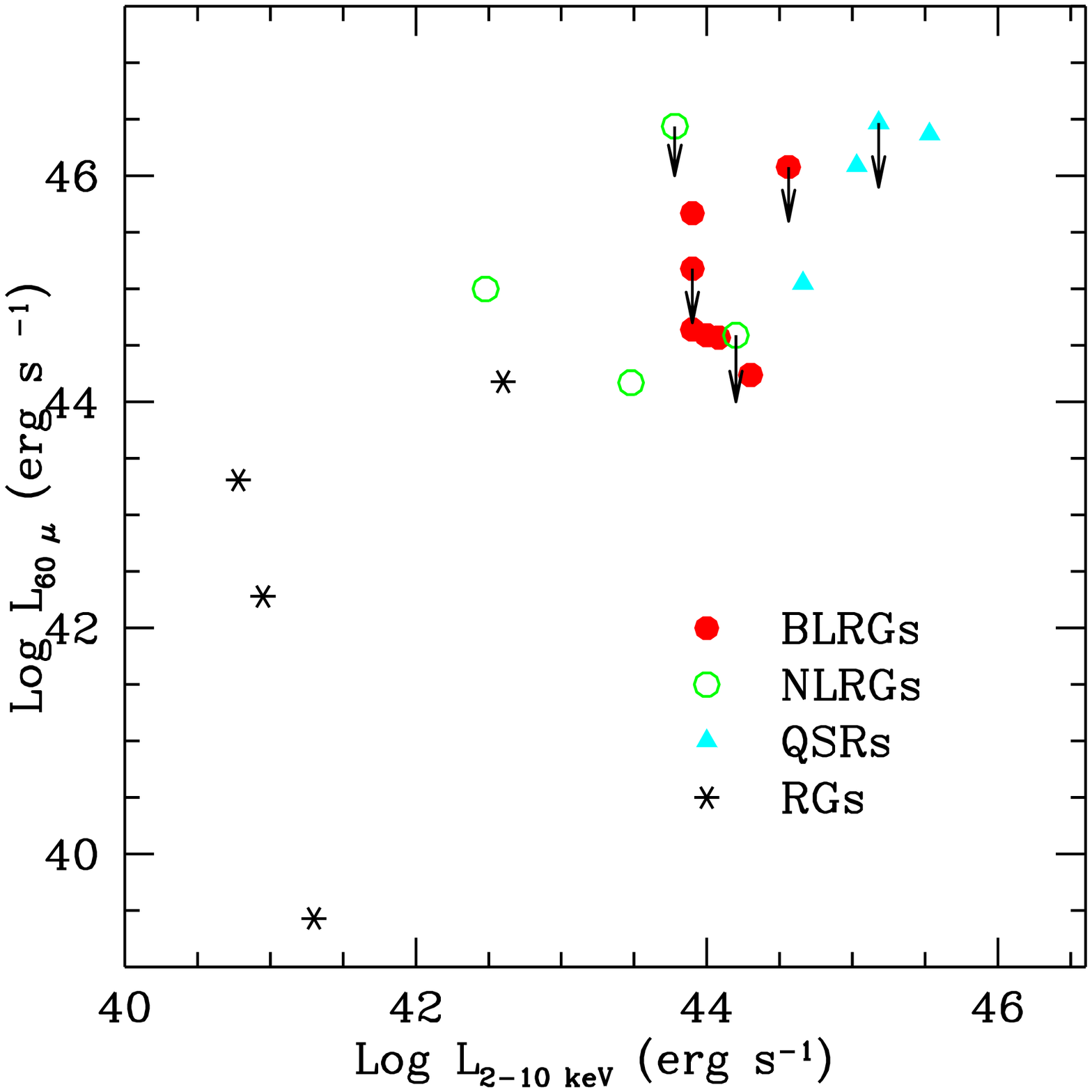,height=3.5in}}
\caption{Correlation of the intrinsic 2--10~keV luminosity with the
[OIII] luminosity {\it (a, Left)}, and the FIR luminosity at 60$\mu$
{\it (b, Right)}. Both correlations are significant after subtracting
the redshift at 99.2\% and 91\% confidence, respectively, from a
survival analysis. }
\end{figure}  

\section{Weak-Line Radio Galaxies} 

Our sample of RGs includes 5 Weak Line Radio Galaxies (WLRGs),
powerful FR~II sources which are underluminous in [O{\,\sc iii}]
emission (by a factor 10) with respect to other radio galaxies with
similar radio power. WLRGs are also characterized by a low ionization
parameter in their narrow-line regions, as indicated by the [O{\,\sc
ii}]/[O{\,\sc iii}] ratio (Tadhunter et al. 1998). One central
question is the cause of such low [O{\,\sc iii}] luminosities, which
could be a weak, hidden AGN or intrinsically different properties of
the line-emitting gas (Tadhunter et al. 1998).

The {\it ASCA} spectra of WLRGs can be modeled with a Raymond-Smith
plasma ($kT\sim1$~keV) at soft energies, plus (in 4/5 cases) a power
law component above 2~keV with average photon index $\langle \Gamma
\rangle=1.5$ (dispersion $\sigma=0.30$) and heavy ($N_{\rm H} \sim
10^{21-23}$ cm$^{-2}$) absorption. The intrinsic
(absorption-corrected) 2--10~keV luminosities are in the range $L_{\rm
2-10~keV} \sim 10^{40-42}$ erg s$^{-1}$.  Moreover, a narrow
unresolved Fe K$\alpha$ line is marginally detected ($\lesssim$ 95\%
confidence) in 2 bright WLRGs, with $EW\sim 250$~eV. WLRGs fall at the
faint end of the $L_{\rm X} - L_{\rm [O{\,\sc iii}]}$ diagram of
Figure~2$a$. While a non-AGN origin of the {\it ASCA} power law
component can not be ruled out {\it a priori}, the correlation of
$L_{\rm X}$ with $L_{\rm [O{\,\sc iii}]}$ is suggestive of a hidden
weak AGN in WLRGs (Tadhunter et al.  1998). On the other hand, it is
interesting that the [O{\,\sc ii}]/[O{\,\sc iii}] line ratios of WLRGs
are reminiscent of LINERs (e.g., Filippenko 1996), raising the
possibility that they could represent the radio-loud counterparts of
these objects. Viewed from this perspective, the true nature of WLRGs
may not be easy to ascertain; they may constitute a heterogeneous
population with starburst activity being responsible for the observed
characteristics of some of them. Future X-ray and multiwavelength
observations of WLRGs will help clarifying their true nature.

\section{Discussion}

\noindent{\bf 1. Constraints for unification models.}  Current
unification scenarios for RL AGNs postulate that orientation effects
are responsible for their optical classification (e.g., Urry \&
Padovani 1995). In particular, NLRGs are thought to be observed at
large inclination angles relative to their jet axis. Hence their
central engines, intrinsically similar to BLRGs/QSRs, would be
completely obscured by an opaque torus. Our {\it ASCA} results are
qualitatively in agreement with the unified models. The continuum
slope distributions of the three classes are similar (Figure 1$a$),
and the hard power law in NLRGs is absorbed by column densities
spanning the large range 10$^{21-23}$ cm$^{-2}$.  However, it is
interesting that the {\it intrinsic} 2--10~keV luminosities of RGs are
significantly lower (by about an order of magnitude) than those of
other types of RL AGNs (Figure~1$b$). This suggests that not all
``type-2'' AGNs contain luminous quasars in their cores. It is also
puzzling that we detect significant cold absorption, in excess of the
Galactic value, in a few BLRGs and QSRs, with column densities similar
to those found NLRGs, while the line of sight to ``type-1'' objects
should be unobstructed.  Interestingly, the BLRGs and QSRs with excess
cold absorption are also powerful FIR emitters, suggesting that the
high columns detected with {\it ASCA} could be associated to dust
responsible for re-radiating the power at lower frequencies.
 
\noindent{\bf 2. Radio-loud vs. radio-quiet AGNs.} We compared the
distribution of the 2--10~keV photon indices of BLRGs with the
distribution of indices for radio-quiet Seyfert 1s studied by Nandra
et al. (1997), in a matched range of intrinsic X-ray luminosities (6
objects). A Kolmogorov-Smirnov test shows the two distributions are
not demonstrably different, with a probability that they are different
of 85\%. However, the mean values, $\langle \Gamma_{\rm BLRG}
\rangle=1.65$ ($\sigma_{\rm BLRG}=0.27$) and $\langle \Gamma_{\rm
Sy\,1} \rangle=1.88$ ($\sigma_{\rm Sy\,1}=0.15$), are different at
92\% confidence from a Student t-test.  Thus, there is only marginal
evidence that RL AGNs have flatter hard X-ray slopes than RQ AGNs in
the ASCA energy band. Differences in the accretion flows of the two
AGN types are also present at higher energies ($>$ 9 keV):
our {\it RXTE} observations of BLRGs show them to lack the
spectroscopic signature of Compton reflection (Eracleous \& Sambruna
1999, in prep.; see also Wo\'zniak et al. 1998), that is a hallmark of
Seyfert 1s (Nandra et al. 1997), suggesting that the structure of the
accretion flow in the two AGN types is different.
 
The {\it ASCA} data provide evidence for different conditions of the
gas in the immediate vicinity of the central engine in RL AGNs: while
ionized absorption is common in Seyfert 1s (Reynolds 1997), only 1
BLRG in our sample shows evidence for a warm absorber. The observed
optical depth implies a column density for the ionized gas $N_{\rm
H}^{\rm warm} \sim 1 \times 10^{21}$ cm$^{-2}$, at the lower end of
the distribution for Seyfert 1s of similar X-ray luminosity (George et
al. 1998). This implies either that the density of warm gas along the
line of sight to the black hole in RL AGNs is lower than in Seyferts,
or that the gas distribution is different, e.g., it may be arranged in
a flat disk perpendicular to the radio axis, as suggested by Wills \&
Browne (1986), and many other authors thereafter.

\noindent{\bf 3. Potential contributions of XMM to the study of RL
AGNs.}  Because RL AGNs are fainter in X-rays and intrinsically rarer
than their RQ counterparts, they are underepresented so far in
existing data archives.  The situation is bound to change with {\it
XMM}.  Thanks to the large effective area of EPIC at 6~keV, the
Fe~K$\alpha$ line will be detected and studied in the fainter objects,
including WLRGs, allowing us to explore trends with intrinsic
luminosity.  Moreover, the line profiles of the brighter objects can
be studied in detail using modest exposure times, in order to separate
broad and narrow components and pin down the line production site.  At
lower energies, the RGS will be able to detect narrow absorption
features, easily missed by {\it ASCA}. In addition, the soft excess
observed by {\it ASCA} and earlier instruments can be resolved into
individual lines to $EW$ limits 10 times lower than those attainable
with the {\it AXAF} gratings. These spectra will allow us to perform
accurate plasma diagnostics as well as study the dynamics of the
circumnuclear medium. {\it XMM} observations of RL AGNs will mark
fundamental progress in our understanding of the structure of their
central engines and thus, ultimately, on the nature of the RL/RQ AGN
dichotomy.

\acknowledgements

RMS acknowledges support from NASA contract NAS--38252.

\end{document}